\documentclass[showpacs,aps,preprintnumbers,prb,amsmath,amssymb]{revtex4}
\usepackage{graphicx}

\begin{document}
\title{Critical scaling of the renormalized single-particle wave function near the Mott-Hubbard
transition}

\author{J. Spa\l{}ek$^{1,2}$, J. Kurzyk$^{3}$, R. Podsiad\l{}y$^{1}$,
and W. W\'ojcik$^{3}$}
\affiliation{$^{1}$Marian Smoluchowski Institute of Physics, Jagiellonian University,
Reymonta 4, 30-059 Krak\'ow, Poland \\
$^{2}$Faculty of Physics and Applied Computer Science, AGH University of Science and Technology,
Reymonta 19, 30-059 Krak\'ow, Poland \\
$^{3}$Institute of Physics, Cracow University of Technology, Podchor\c{a}\.zych 1, 30-084 Krak\'{o}w,
Poland}
\begin{abstract}
We present a quantum critical behavior of the renormalized single-particle Wannier function,
calculated in the Gutzwiller correlated state near the insulator-metal
transition (IMT) for cubic lattices. The wave function size and its maximum,
as well as the system energy scale with increasing lattice parameter $R$ as $R^{n}$.
Such scaling is interpreted as the evidence of a dominant role of the Coulomb repulsion.
Relation of the insulator-metal transition lattice-parameter value $R=R_{C}$ to the original
{\em Mott criterion\/} is obtained. The method is tested by comparing our results with the exact
approach for the Hubbard chain.
\end{abstract}
\pacs{70}
\maketitle

Insulator-metal transition (IMT) of the Mott-Hubbard type represents one of the central problems in quantum
physics, since it exemplifies a transition from well defined atomic states to delocalized
(Bloch- or Fermi-liquid-type) states in a solid\cite{1} and in other systems.\cite{2} The model system
is a
lattice of spin-1/2 fermions with one particle per atomic site. Mott expressed this localized-delocalized
transformation in terms of critical density $n_{C}$ of fermions (or equivalently, in terms of critical
interparticle distance $R_{C}=n_{C}^{-1/D}$, where $D$ is the system dimensionality) and the effective Bohr-
radius $a_{B}$ of the atomic states formed at the transition, in the form $ n_{C}^{-1/D} a_{B}\simeq
0.25$. On the other
hand, Hubbard\cite{3} and others\cite{4} formulated the criterion in the form that at the transition the
Coulomb repulsive energy among the particles (of magnitude $U$) is equal to their kinetic (band) energy
characterized
by the bare bandwidth $W$, $U\simeq W$. These two criteria are regarded as
equivalent.

The insulator-metal transition is also well defined as a phase transformation in
thermodynamic sense.\cite{5} As close to IMT the band energy (negative) is almost compensated by the repulsive
Coulomb interaction, the much smaller thermal or energy can drive the metallic system towards
the state with localized spins. The resultant phase diagrams contain
first-order pressure-temperature transition line with a terminal critical point
at temperature $T_{cr}>0$.
The existence of the critical point was confirmed experimentally quite recently\cite{6} and
shown to represent properties of the Van der Waals liquid-gas point
in the $D=3$ case of V$_{2}$O$_{3}$:Cr, (see Limelette et al.\cite{6}). 
The basic question remains under what conditions a quantum
critical point (QCP) appears at $T_{cr}=0$ between metallic and insulating phases, as its existence
marks explicitly the fundamental boundary between the atomic and the condensed (delocalized) quantum states.
The appearance of such QCP is usually obscured in real systems by the presence of antiferromagnetism
on (at least) one side of the transition.\cite{7} Nonetheless, the answer to the above question
even without inclusion of magnetic ordering would
delineate the basic characteristics of the (quantum) critical point (at $T=0$), which is of
{\em non-Landau character\/}.\cite{8}

As IMT involves a drastic change of the nature of states individual carriers, the natural
and not addressed so far question is how their wave-function changes when approaching the
transition from either side. The parameterized-model approaches\cite{3,4,5}
leave the Wannier functions determining the parameters as fixed. On the other hand, the original Mott
approach introduces intuitively the concept of an emergent atomic state
{\em at the instability\/} of the metallic state.
Thus the missing question is: how the correlations
and the single-particle aspects are interrelated microscopically?

We provide the explicit description of electronic states close to IMT in which both the
interelectronic correlations and
the single-particle Wannier wave functions $\{ w_{i}({\bf r})\}$
are treated {\em on the same footing\/}. As the result, we introduce a {\em singular\/} behavior
of the single-particle (Wannier) wave function renormalized by the correlations. This nonanalytic
behavior (and associated with it unique scaling laws) are only possible when the wave function is calculated
a posteriori, i.e. in the correlated state, not {\em before\/} the correlations are taken into account.
Such an approach is indispensable in the situation when the single-particle and interaction energies
are comparable or the interaction is even dominant.
In this sense, our analysis complements that performed within
either LDA+DMFT\cite{9} or LDA+U\cite{10} methods, which contain parameters
characterizing electronic correlations introduced after the LDA calculations have been carried out.
We should also underline, that our method was applied earlier to both
correlated nano-\cite{11} and macro-\cite{12}systems and is free from the double counting of
the repulsive Coulomb interaction. Furthermore, the microscopic parameters are calculated explicitly
within the mutually consistent procedure.
Therefore, even though the presented below
results concern model results, they should be regarded as an essential ingredient to be implemented
in modeling real systems near IMT.

We start from extended Hubbard model with inclusion of intersite Coulomb interaction
and ion-ion interaction as represented by the Hamiltonian\cite{12,13}
\begin{equation}
H\,=\,\epsilon_{a}^{eff}\,\sum_{i}\, n_{i}\,+\,
\sum_{i\neq j\sigma}\, t_{ij}\, a_{i\sigma}^{\dagger}\, a_{j\sigma}\,+\,
U\,\sum_{i}\,
n_{i\uparrow}\, n_{i\downarrow}\,
\label{eq:w1}
\end{equation}
$$
+\,\frac{1}{2}\,\sum_{i\neq j}\,
K_{ij}\,\delta n_{i}\,\delta n_{j}\,+\, V_{ion-ion}\,,
$$
where the first term describes the effective atomic energy with
\begin{equation}
\epsilon_{a}^{eff}\,=\,
\epsilon_{a}\,+\, \frac{1}{2\, N}\,\sum_{i\neq j}\,\left(
\, K_{ij}\,+\, 2/\, R_{ij}\,\right)\,,
\label{eq:w2}
\end{equation}
the second represents the hopping between the nearest-neighboring sites, the third
the Hubbard (intraatomic) and the fourth is a part of intersite Coulomb interaction,
with $\delta n_{i}\equiv 1-n_{i}$ being the deviation from the integer electron occupancy
$n_{i}=1$, and $K_{ij}$ - the intersite Coulomb interaction. The term $(2/R_{ij})$
in (\ref{eq:w2}) expresses the classical Coulomb repulsion (in atomic units) for two ions
separated by the distance $R_{ij}$.

In the Mott-Hubbard insulating state the hole-hole correlations are absent i.e. $\langle (1-n_{i})(1-n_{j})\rangle\simeq 0$.
The role of these correlations is indeed neglible if the fundamental correlation function $d^{2}\equiv
\langle n_{i\uparrow}n_{i\downarrow}\rangle$ vanishes for $U\sim W$.

The microscopic parameters of this model are expressed via the Wannier functions
$\{ w_{i}({\bf r})\} \equiv \{ w({\bf r}-{\bf R}_{i})\}$\cite{11,12} as follows:
$\epsilon_{a}\equiv \langle  w_{i}\vert  H_{1}\vert w_{i}\rangle$,
$t_{ij}\equiv \langle  w_{i}\vert  H_{1} \vert w_{j}\rangle$,
$U\equiv \langle  w_{i} w_{i}\vert V_{12} \vert w_{i} w_{i}\rangle$, and
$K_{ij}\equiv \langle w_{i} w_{j}\vert  V_{12} \vert w_{i} w_{j}\rangle$,
where $H_{1}$ is the Hamiltonian for a single particle in the system, and $V_{12}$
represents interparticle (Coulomb) repulsion. The Wannier functions are expressed
in terms of adjustable Slater functions, i.e.
$w_{i}({\bf r})=\beta \Psi_{i}({\bf r})-\gamma \sum_{j=1}^{z} \Psi_{j}({\bf r})$
where $z$ is the number of nearest neighbors, $\beta$ and $\gamma$ are mixing coefficients,
and $\Psi_{i}({\bf r})\equiv (\alpha^{3}/\pi )^{1/2}\exp (-\alpha\vert {\bf r}-{\bf R}_{i}\vert)$
is the $1s$ Slater function centered on the site $i\equiv {\bf R}_{i}$.
In the concrete calculations, they are represented by adjustable Gaussians
(e.g. STO-7G basis in dimensions $D=1,2,\mbox{ and }3$) and where discussed before.\cite{11,12}

As said above, the fundamental principle behind our approach is that the wave functions
$\{ w_{i}({\bf r})\}$ are treated on the same footing as the diagonalization of (\ref{eq:w1})
in the Fock space.
Such diagonalization is possible in an exact manner for nanoscopic and
infinite Hubbard chain $(D=1)$ systems.\cite{11-13} However, also the Gutzwiller wave-function
(GWF) and the Gutzwiller-ansatz (GA) approximations lead to close results in the latter case (see below)
provided the wave functions $\{ w_{i}({\bf r})\}$ are properly readjusted variationally in the correlated
state to achieve the ground-state energy as a global minimum for a given lattice parameter $R$.
In other words, the energy of the correlated state is readjusted
iteratively multiple times by adjusting the wave function size $\alpha^{-1}$ and in effect,
also the microscopic parameters. As a result of such
iterative approach, we obtain the renormalized wave function
and the microscopic parameters, as well as the ground-state energy, and most importantly, scaling of the physical
properties, all as a function of $R$. Below we analyze first the results obtained for GA for three-dimensional
cubic lattices, before testing their validity for $D=1$ and $2$ situations. The main emphasis is lead
on the novel singular scaling properties near the Mott-Hubbard critical spacing $R=R_{C}$.

\begin{figure}
\includegraphics[width=0.45\textwidth]{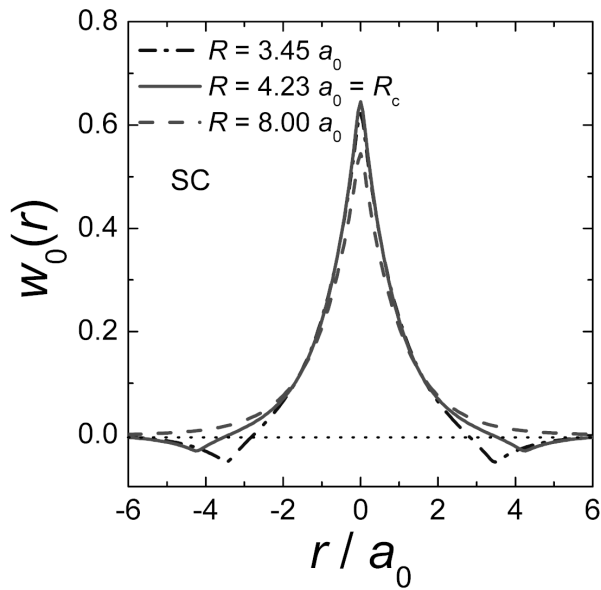}
\caption{\label{fig:Fig1}(Color online) Exemplary shapes of the Wannier function centered at the
site ${\bf R}_{i}=0$ for simple cubic (SC) lattice and for three lattice spacings marked:
$R<R_{C}$ (dot-dashed line), $R=R_{C}$ (solid line), and $R>R_{C}$ (dashed line). $a_{0}$
is the $1s$ Bohr radius.}
\end{figure}

In Fig. 1 we plot the exemplary Wannier function centered at ${\bf R}_{i}=0$ for simple cubic
(SC) lattice for $R<R_{C}$ (dot-dashed line) $R=R_{C}$ (solid line), and $R\gg R_{C}$
(dashed line). One observes immediately that the wave function for $R>R_{C}$ is more extended
than that for $R=R_{C}$ and with increasing $R$ gradually reaches its atomic value $a_{0}$.
To study this nontrivial effect in detail, we have plotted in Fig. 2 the
relative inverse size $\delta \alpha/\alpha\equiv \vert\alpha(R)-\alpha(R_{C})\vert /\alpha(R_{C})$ as a function
of relative lattice spacing $\delta R/R\equiv (R-R_{C})/R_{C}$ in the regime $R>R_{C}$. A clear 
$d\alpha/dR$ discontinuity (cf. inset) is observed at $R=R_{C}$ for all cubic lattices, a rather unique
and unexpected feature, which is completely absent for the case with bare (unrenormalized) wave functions.
Note that the wave function is the narrowest at the Mott-Hubbard transition at which $d^{2}\approx 0$
($d^{2}>0$ in the metallic phase). To determine the universality of the behavior we have replotted
in Fig. 3 the result of Fig. 2 for $R>R_{C}$ in a doubly logarithmic scale. One observes a clear
power law scaling $\delta\alpha/\alpha\sim [(R-R_{C})/R_{C}]^{{\it s}}$, with ${\it s}= 0.96\pm 0.1$
for not too large $R$.
The results for $R<R_{C}$ exhibit also a similar scaling property
for $R\lesssim R_{C}$, with the exponent ${\it s}$
slightly lower $({\it s}\simeq 0.93)$ and weakly dependent on the type of cubic lattice selected. This means
that the results in the metallic phase show a lesser degree of universality.
Note that this power-law scaling describes the singular behavior of the wave-function size
$a_{B}= 1/\alpha$.

\begin{figure}
\includegraphics[width=0.45\textwidth]{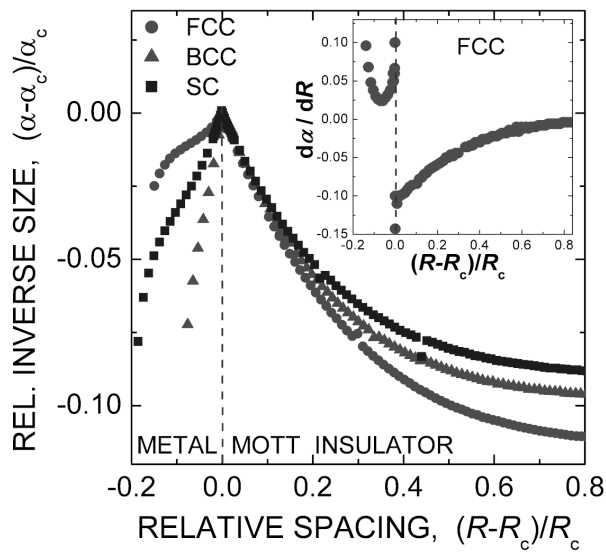}
\caption{\label{fig:Fig2} (Color online) Relative inverse wave function size
$\delta\alpha/\alpha$ as a function of relative lattice parameter $\delta R/R$
for simple cubic (SC), body centered cubic (BCC), and face centered cubic(FCC) lattices.
The Mott-Hubbard transition is marked by the vertical dashed line. Inset: derivative
$d\alpha/dR$ vs. $\delta R/R$ with a singularity at $\delta R\equiv R-R_{C}=0$.}
\end{figure}

Lattice dependence of the scaling with $\delta R/R$ is also observed for the ground
state energy, as shown in Fig. 4. The approximate dependence $\sim R^{t}$ of $E_{G}$ when approaching
the atomic limit can be attributed to the dominant role of the Coulomb repulsion between
the electrons localized on neighboring ions.

Both the Figs. 2 and 4 demonstrate the appearance of a simple scaling properties of $R^{\pm 1}$ type.
This dependence could be seen in the pressure dependence of the orbital size when close to the critical value
$R=R_{C}$. However, the relative changes of $\alpha$ (cf. Fig. 2) are rather subtle and the question
is if they can be observable in the present day e.g. scanning tunneling observations of atomic
orbitals. For that purpose we have plotted in Fig. 5 the maximal value $w_{0}(0)$
of the Wannier function relative to that at $R=R_{C}$. The occupancy depletion reaches up to
$20\%$ of the peak value upon change of $20-25\%$ of the lattice constant. Also, one observes
the behavior of the same type, as that for $\alpha$ in Fig. 2. However, a clear scaling takes place only for $R>R_{C}$
(see inset), where it assumes the shape $\sim (R/R_{C})^{{\it u}}$, with ${\it u}=0.92\pm 0.01$,
not much different from that displayed in Fig. 3. On the basis of the results one can say that our approach
provides evidence for long-range effects in the single-particle wave function which are pronounced
already in the Mott-Hubbard insulating state. In other words, the Hubbard split subband
picture\cite{3} of the Mott insulator is concomitant with a strong renormalization of the wave
function characteristics near the transition point.

\begin{figure}
\includegraphics[width=0.45\textwidth]{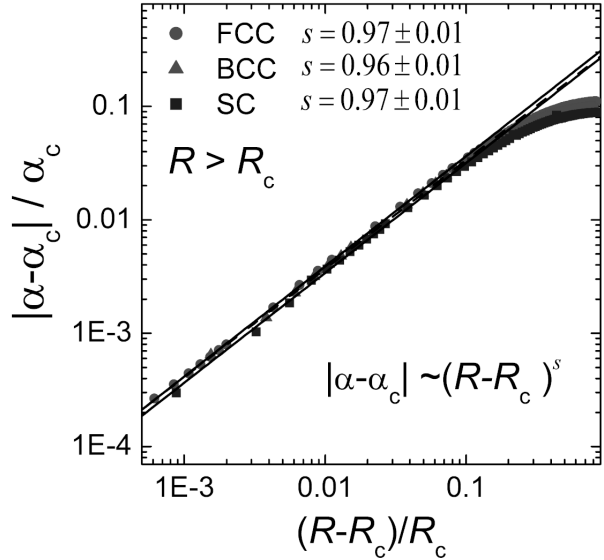}
\caption{\label{fig:Fig3} (Color online) Detailed scaling $\delta \alpha/\alpha$ vs.
$\delta R/R$ in the doubly logarithmic scale. The straight lines represent the fitted
curves $\sim (R/R_{C})^{{\it s}}$.}
\end{figure}

\begin{figure}
\includegraphics[width=0.45\textwidth]{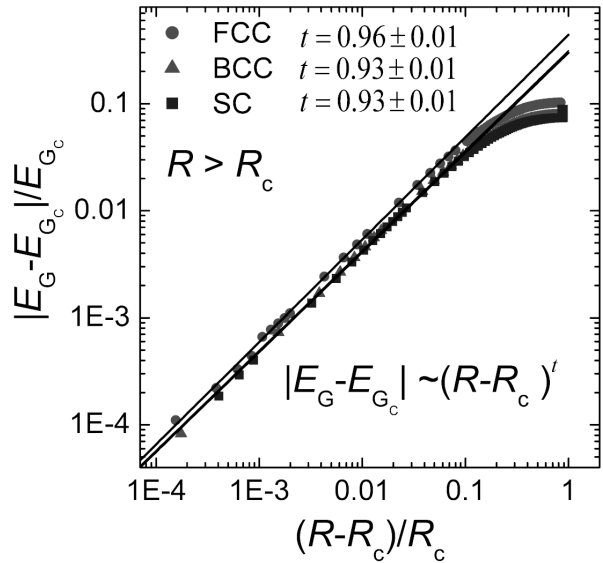}
\caption{\label{fig:Fig4} (Color online) Same scaling as in Fig. 3 but for the ground state
energy; the fitted straight-line characteristics are listed.}
\end{figure}

Finally, we have also examined the role of lattice dimensionality on the singular behavior of the
wave function. For that purpose we plot in Fig. 6 the inverse wave-function size vs. $R/a_{0}$
for the linear chain within the modified exact Lieb-Wu (LW),\cite{15} Gutzwiller-
wave-function (GWF),\cite{16} and Gutzwiller-ansatz (GA)\cite{12,13,17} solutions.
None of the methods provides in $D=1$ case (and GA for $D=2$) the type of singular behavior observed in Figs. 2 and 4
for $D=3$ case.
What is more important, all three methods provide quite similar results for $D=1$. This circumstance
gives us with some confidence, that the modified above GA method for $D=3$ cubic lattice bears
physical relevance to the problem considered as it removes, among others, a spurious IMT transition
for $D=1$ lattice. Therefore, it appears that only correlation effect treated in GA combined with the singular
behavior of the wave function provide a correct IMT characterization.
Furthermore, in the inset to Fig. 6 we plot
the Wannier function maximum vs. $R/a_{0}$ for linear chain (CH), square (SQ), and
triangular lattices (TR), respectively. The results for $D=2$ are obtained for GA solution.
Again, no critical behavior is observed. On the basis of these results we draw the conclusion
that within our approach the critical behavior appears only for $D=3$.
The striking coincidence of these results to the critical dimensionality $(D>2)$ for the
onset of Landau-Fermi liquid stability should therefore be mentioned.\cite{18}

\begin{figure}
\includegraphics[width=0.45\textwidth]{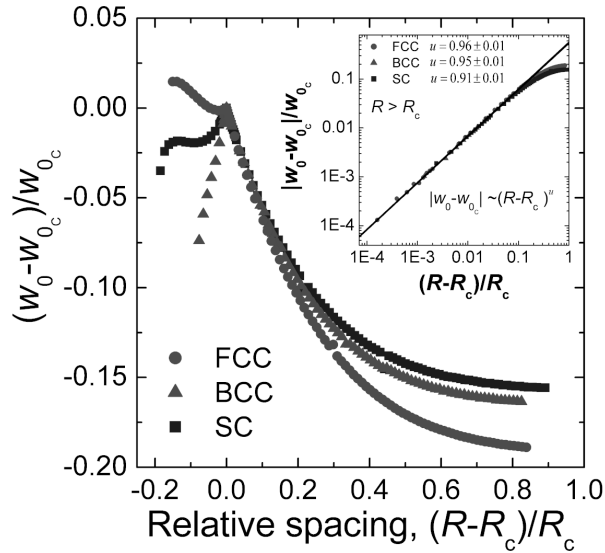}
\caption{\label{fig:Fig5} (Color online) Same as in Fig. 2, but for the maximum $w_{0}(0,0,0)$
of the Wannier function. Inset: Detailed scaling of the Wannier function maximum
(of same type as in Fig. 3). Fitted straight line represents the function $\sim (R/R_{C})^{0.92}$.}
\end{figure}

\begin{figure}
\includegraphics[width=0.45\textwidth]{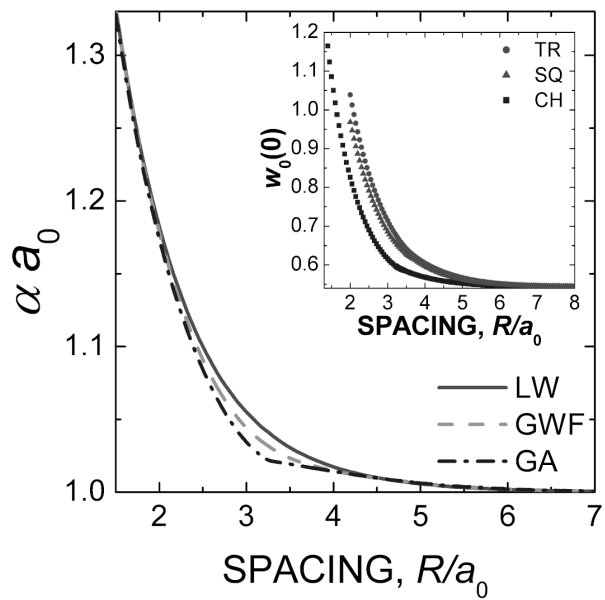}
\caption{\label{fig:Fig6} (Color online) Inverse size of the renormalized wave function
for linear chain (CH) vs. $R/a_{0}$ within the exact Lieb-Wu (LW), the Gutzwiller-wave-function
(GWF), and the Gutzwiller-ansatz (GA) solutions. No singular behavior is observed at any $R$.
Inset: Wannier-function maximum $ w_{0}(0)$ vs. $R/a_{0}$ for linear chain (CH),
square (SQ), and triangular lattices (TR), with no discontinuity detected in either of the cases.}
\end{figure}

One additional basic feature of our approach should be mentioned. Namely, in Table I
we list microscopic characteristic of the Mott-Hubbard transition. These results intercorrelate nicely
with the Mott criterion in the following manner. The carrier concentration is $n_{C}=1/R_{C}^{3}$
for SC, $2/R_{C}^{3}$ for BCC, and $4/R_{C}^{3}$ for FCC. Thus $n_{C}=n/R_{C}^{3}$,
with $n=1,2,\mbox{ and }4$, respectively. Therefore, the Mott criterion takes the form:
$n_{C}^{1/3}\cdot a_{B}= n^{1/3}/(R_{C}\alpha_{C})=0.20$, $0.26$, and $0.33$,
for SC, BCC, and FCC lattices, very close values to that obtained originally by Mott.\cite{1} 
This connection provides
additional argument of our approach.

\begin{table}
\caption{Microscopic parameters at the critical point for GA solutions for the lattices specified.}
\begin{tabular}{c|ccc}
\hline\hline
Struct. & $\alpha /a_{0}$ & $R_{C}/a_{0}$ & $(U/W)_{C}$ \\
\hline
SC & 1.099 & 4.236 & 1.337 \\
BCC & 1.109 & 4.384 & 1.080 \\
FCC & 1.128 & 4.351 & 0.880
\end{tabular}
\end{table}

In summary, we have extended the standard treatment of the Mott-Hubbard transition 
by incorporating into
the consistent scheme the wave function renormalization and demonstrating its
singular behavior.
On the basis of our results one can expect also important effect of temperature and
magnetic ordering
onto this behavior. 
The long-range effects on the wave function shape
develop already in the insulating phase and become the strongest at the transition by narrowing
the atomic wave-function down by about $10\%$. The wave-function characteristic length $\alpha^{-1}$
plays the role of a coherence length for a single fermion in the sea of all other particles.
Such an approach can also be extended to orbitally degenerate $3d$ along the lines
proposed earlier.\cite{19,20}

The work was supported by the Grant No. NN 202 128 736 from Ministry of Science
and Higher Education. Also, the research is performed under the auspices of the
ESF Network INTELBIOMAT.


\begin{thebibliography}{99}
\bibitem{8}
This means, there is no symmetry change
associated with transition at that point. Hence its non-Landau character, as in the case for V$_{2}$O$_{3}$:Cr.
\end{thebibliography}
\end{document}